\documentclass[preprint,secnumarabic,amssymb, aps, prd]{revtex4}

\usepackage{amsmath}
\usepackage{mathtools}
\usepackage{graphicx}
\usepackage{epsfig}
\usepackage{latexsym,oldgerm}
\usepackage{longtable}
\def\beq{\begin{equation}}
\def\eeq{\end{equation}}
\def\bea{\begin{eqnarray}}
\def\eea{\end{eqnarray}}
\def\nn{\nonumber}
\begin{document}
\title{Active Brownian motion of an asymmetric rigid particle}
\author{Gulmammad Mammadov}
\email[Email: ]{gmammado@syr.edu}
\affiliation{Department of Physics, Syracuse University, Syracuse, NY 13244-1130, USA}


\begin{abstract}
Individual movements of a rod-like self-propelled particle on a flat substrate are quantified. Biological systems that fit into this description may be the Gram-negative delta-proteobacterium \emph{Myxococcus xanthus}, Gram-negative bacterium \emph{Escherichia coli}, and \textit{Mitochondria}. There are also non-living analogues such as vibrated polar granulates and self-driven anisotropic colloidal particles. For that we study the Brownian motion of an asymmetric rod-like rigid particle self-propelled at a fixed speed along its long axis in two dimensions. The motion of such a particle in a uniform external potential field is also considered. The theoretical model presented here is anticipated to better describe individual cell motion as well as intracellular transport in 2D than previous models.

\end{abstract}
\maketitle

\section{Introduction}
In classical physics the Brownian motion is the erratic motion of microscopic particles due to non-zero value of net stocastic forces at any given time exerted on them by atoms or moleculs of the surrounding medium. Such microscopic particles don't play active role in the motion and  their irreversible dissipation of energy is compensated by reversible thermal fluctuations as expressed in the fluctuation-dissipation theorem. 

In contrast, if the particles have internal or
external energy source leading to self-propulsion, their motion will have directional biasness. Such motion is reffered to as the \textit{active} Brownian motion. Some example subjects of active Brownian motion relevant to our study are self-driven colloidal
particles, living microorganisms, and vibrated \textit{Janus} particles~\cite{sven, arshad, bala, andreas, liverpool}. These systems exhibit many interesting  
nonequilibrium effects understanding of which have been the research focus of many scientists in recent years. 

In this context, a phenomenological model of active Brownian motion has been developed
based on principle of energy take up, store and dissipation into
kinetic degrees of freedom~\cite{ebeling0, ebeling}. Other research team, J. R. Howse
\textit{et al.}, experimentally studied active Brownian motion of
spherical polystyrene particles that are self-propelled due to
asymmetric surface activity~\cite{golestanian}. They interpreted the data using an
expression for the mean-squared displacement that takes into account
rotational diffusion of the self-propulsion velocity. More recently motion of
deformable self-propelled particles analytically has been studied
assuming that the particles can deform from circular shape when
the propagation velocity is increased~\cite{taka}. It was shown that
such a particle undergo a bifurcation from straight motion to a
circular motion by increasing the propagation velocity. Also,
Brownian dynamics of a microswimmer composed of three spheres which
propels itself in a viscous fluid by spinning motion of the spheres
under zero net torque has been studied~\cite{vlad}. Finally, in
ref~\cite{sven} dynamics of a Brownian circle swimmer has been
studied when the driving force does not coincide with the
propagation direction. It was hypothesized that when the self-propulsion
force and the particle orientation are in line, the motion would be
along a straight line just perturbed by random fluctuations.

In this paper, we present detailed study of the active Brownian
motion of an asymmetric particle when the self-propulsion force
and the particle orientation are in line. Unlike previous studies our
approach takes into account the particle asymmetry.

Collective behavior of such rods, which interact with each other through excluded volume, has
been theoretically studied~\cite{apa} and it was shown that the
self-propulsion enhances the longitudinal diffusion coefficient. Our
calculations, however, show that in the case of one rod both the
longitudinal and the transverse diffusion coefficients are enhanced
and assume the same value at long times.

The paper is organized as following: After introducing the model, we start by calculating the mean squared displacements and corresponding diffusion tensors for two cases: when a rod is self-propelled at constant speed and when such a self-propelled rod is in external constant potential field. The paper is concluded by the disucssion of the results.

\section{The model}
\begin{figure}[t!]
  \begin{center}
    \includegraphics[width=0.4\textwidth]{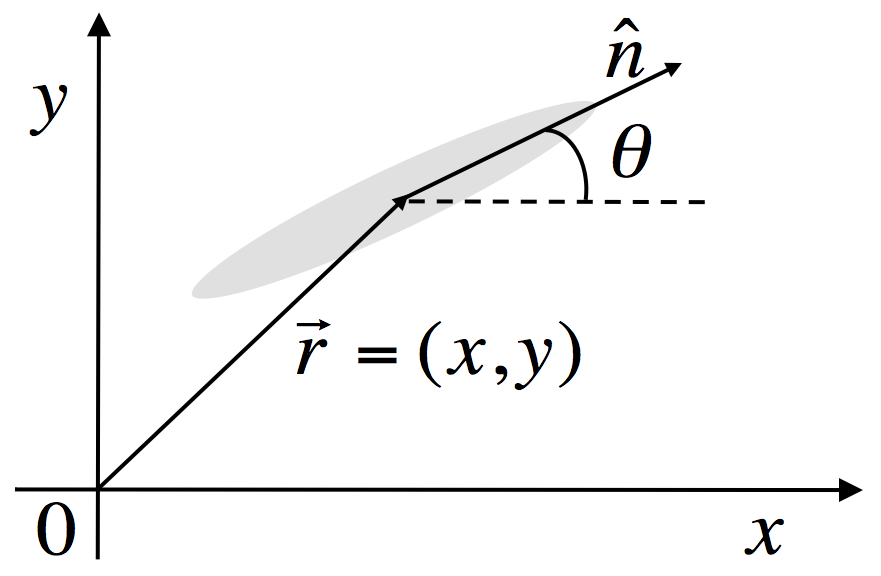}
    \caption{The diagram shows an ellipsoidal rigid particle self-propelled along its long axis.\label{fig:0}}
  \end{center}
\end{figure}

The model consists of an
ellipsoidal rigid particle that is self-propelled along its long
axis and undergoing Brownian motion in two dimensions~(see
FIG.~\ref{fig:0}). Note that this model can be generalize to higher
dimensions using the Pythagorean theorem. 

At a given time $t$ the rod can be described by the position vector
of its center of mass $\textbf{r}(t)$ and the angle $\theta(t)$
between the $x$ axis of the \textit{lab-frame} and the long axis of
the rod. In this frame the self-propulsion speed, which is taken along
the long axis of the rod, is given by
 $v\hat{\textbf{n}}(t)$, where $\hat{\textbf{n}}(t)\equiv
(\cos\theta(t),\sin\theta(t))$ is a unit vector along the long axis
of the rod. 
In the presence of external force  $\textbf{F}$ and torque $\tau $ the dynamics of the rod is 
governed by the coupled Langevin equations, given by
\beq\label{1}
\partial_{t}r_i=\Gamma_{ij}(\hat{n})F_j+v\hat{n}_i(t)+\xi_i(t),\eeq
\beq\label{2}
\partial_t\theta=\Gamma_R \tau+\xi_R (t),\eeq
where $i=x, y$ for $2D$ and
$\Gamma_{ij}$, $\Gamma_R$ are the mobilities relating velocity and
angular velocity to force and torque, respectively.
In the  \textit{body-frame} denoting the mobility along the long axis of the rod by
$\Gamma_\|$ and perpendicular by $\Gamma_\bot$, the mobility
tensor in the lab-frame takes the following form
\beq\label{3}
\Gamma_{ij}(\hat{n})=\Gamma\delta_{ij}+\frac{\Delta\Gamma}{2}M_{ij}(\theta),
\eeq where $M_{ij}(\theta)=\left(
                             \begin{array}{cc}
                               \cos2\theta & \sin2\theta \\
                               \sin2\theta & -\cos2\theta \\
                             \end{array}
                           \right)
$, $\Gamma=\frac{\Gamma_\|+\Gamma_\bot}{2}$, and
$\Delta\Gamma=\Gamma_\|-\Gamma_\bot$.
The rotational noise $\xi_R$ is a Gaussian random variable with zero mean and variance 
\beq\label{4}
\langle \xi_R(t) \xi_R(t')\rangle=2k_B T \Gamma_R\delta(t-t'),
\eeq
whereas $\xi_x(t)$ and $\xi_y(t)$ are Gaussian at a fixed $\theta(t)$ with zero mean and variances depending on orientation at a given time
\beq\label{5}
\langle\xi_i(t) \xi_j(t')\rangle_{\theta(t)}=2k_B T\Gamma_{ij}(\theta(t))\delta(t-t'),
\eeq
where $k_B$ is the Boltzmann constant and $T$ is
the effective temperature. The ensemble
average is calculated using
$\langle\cdots{}\rangle=\frac{1}{2\pi}\int_0^{2\pi}\langle\cdots{}\rangle_{\theta_0}d\theta_0$
where $\langle \cdots{}\rangle_{\theta_0}$
indicates average both over $\xi_i(t)$ and $\xi_R(t)$ at a fixed initial
angle.

The self-propulsion force doesn't induce torque, which in turn enables the rod to rotate freely about the axis passing through its center of mass and perpendicular to the plane of motion. This means that the angular displacements obey the Gaussian statistics with constant rotational
diffusion coefficient. In the lab-frame the shape asymmetry of the rod makes the translational and rotational displacements coupled, which complicates the analytic examination of the motion. The Gaussianity of the angular displacements, however, simplifies the matter and leads to detailed analytic results. 

\section{Self-propulsion at constant speed}

As we mentioned previously, to carry out the calculations analytically we shall set $\tau=0$. Therefore in the absence of external torque integration of Eq.~\eqref{2} gives
$\Delta\theta(t)=\int_0^{t}\xi_R(t')dt'$. Then with the aid of Eq.~\eqref{4} for
the mean-squared angular displacement (the second moment) we obtain
$$\langle[\Delta\theta(t)]^2\rangle=\int_0^{t}\int_0^{t}\langle\xi_R(t')\xi_R(t'')\rangle
dt'dt''=2D_R t.$$

Subsequantly the noise avarages of the sinusoidal functions will be evaluated using the following identity~\cite{schulten}:
\beq\label{6}
\langle e^{i[s_1\Delta\theta(t_1)+
s_2\Delta\theta(t_2)]}\rangle_{\theta_0}=e^{-D_R[s_1^2t_1+s_2^2t_2+
2s_1s_2\min(t_1,t_2)]},\eeq
where $D_R=k_BT\Gamma_R$ is the rotational diffusion coefficent.

Following the derivation of Ref.~\cite{han} in the absence of external force $\textbf{F}$ one can integrate Eq.~\eqref{1} with respect to time and use
Eq.~\eqref{a6} to calculate the ensemble average of the sinusoidal
functions. Then for the mean displacement one obtains
\beq\label{7}
\langle\Delta\textbf{r}\rangle_{\theta_0}=v\tau_1(\cos\theta_0, \sin\theta_0),
\eeq
where $\tau_n=\frac{1-e^{-nD_Rt}}{nD_R}$ and $\tau_R=1/2D_R$ is
the rotational time required for one radian diffusion.
Finally, using Eq.~\eqref{7} for the magnitude of the mean
displacement of the rod one has 
\beq\label{8}
|\langle \Delta\textbf{r}\rangle_{\theta_0}|=[\langle\Delta
x\rangle_{\theta_0}^2+\langle\Delta
y\rangle_{\theta_0}^2]^{\frac{1}{2}}=2v\tau_R[1-e^{-\frac{t}{2\tau_R}}].
\eeq

To find the mean-squared displacement tensor resulting from
the integration of Eq.~\eqref{1} with $\textbf{F}=0$, we note that on averaging over the
noise, cross terms such as $\langle\xi_i(t_1)n_j(t_2)\rangle$ vanish
since they are first order in the noise. After doing so, in the
expression for the mean-squared displacement tensor we are left with
only two terms
\bea\label{9}
\langle\Delta x_i(t)\Delta
x_j(t)\rangle_{\theta_0}= \int_0^{t}d t_1\int_0^{t}d t_2\langle
\xi_i(t_1)\xi_j(t_2)\rangle_{\theta_0} \nn\eea \bea\label{6}&&+
v^2\int_0^{t}d t_1\int_0^{t}d t_2\langle n_i(t_1)
n_j(t_2)\rangle_{\theta_0}. \eea

The first term in Eq.~\eqref{9} has been calculated in ref.~\cite{han} and the second term is given by
Eq.~\eqref{a7}. If we put them together, for the mean-squared
displacement tensor at a fixed initial orientation we obtain
\bea\label{10}
\langle\Delta x_i\Delta x_j\rangle_{\theta_0}=[2Dt+2\tau_R
v^2(t-\tau_1)]\delta_{ij}\nn\\+[\Delta D\tau_4+\frac{2\tau_R
v^2}{3}(\tau_1-\tau_4)]M_{ij}(\theta_0), \eea
and the mean-squared displacement
\bea\label{11}
\langle \Delta\textbf{r}^2\rangle=4Dt+4\tau_R v^2[t-2\tau_R (1-e^{-t/2\tau_R})],
\eea
with $D=k_BT\Gamma$ and $\Delta D=k_BT\Delta\Gamma$.
For the displacement diffusion tensor of an active rod Eq.~\eqref{10}  yields
\bea\label{12}
D_{ij}(t,\theta_0)=\frac{\langle\Delta x_i\Delta
x_j\rangle_{\theta_0}}{2t}=[D_S-\tau_R
v^2\frac{\tau_1}{t}]\delta_{ij}\nn\\+\Delta
D\frac{\tau_4}{2t}[1+\frac{2\tau_R
v^2}{3\Delta D}\frac{\tau_1-\tau_4}{\tau_4}]M_{ij}(\theta_0), \eea
where $D_S=D+\tau_Rv^2$. Averaging Eq.~\eqref{12} over the initial angle,
$\theta_0$, yields a time-dependent displacement diffusion tensor,
\beq\label{13}
D_{ij}(t)=\frac{1}{2\pi}\int_0^{2\pi}d\theta_0D_{ij}(t,\theta_0)=[D_S-\tau_R
v^2\frac{\tau_1}{t}]\delta_{ij}. \eeq 

This shows that the system of non-interacting
self-propelled rods is no longer diffusive as the displacement
diffusion tensor is time dependent. However, in all cases the time
dependence becomes negligible for $t\gg\tau_R$, and as $t$ tends to
infinity unlike the \textit{passive} ellipsoidal rigid
particle case, for the self-propelled ellipsoid the single long-time
diffusion coefficient is enhanced according to
\bea\label{14}
D_{ij}(t)\rightarrow D_S\delta_{ij}. \eea 

When $t\gg\tau_R$ the directional memory is lost and the rod behaves as a perfect
spherical Brownian particle with enhanced diffusion coefficient.

On the other hand, when $t\ll\tau_R$ the diffusion tensor and the mean-squared
displacement are give by  $D_{ij}(t)\simeq [D+\frac{v^2t}{4}]\delta_{ij}$ and $ \langle \Delta
\textbf{r}^2\rangle\simeq4Dt+ v^2t^2$, respectively, indicating ballistic behavior at short times.

\section{Self-propulsion in external constant force potential field}

In the previous section we studied the active Brownian motion of a rigid
rod in the absence of external field and torque. Here we discuss the motion
under the infulance of external constant force. As in the previous case
we assume the rod is constrained to move in a plane. An example of such a system may be active colloidal suspension of asymmetric particles under gravity.

Without loosing any generality, we can always rotate the coordinate system (lab-frame) such that the
external force is along the $x$ axis, $\textbf{F}=F\hat{x}$. Then
using Eq.~\eqref{3} from Eq.~\eqref{1} we can write
\bea\label{15}
\Delta x(t)=F\Gamma t+\frac{1}{2}F\Delta\Gamma\int_0^t\cos2\theta
dt'\nn\\+\int_0^t\xi_xdt'+v\int_0^t\cos\theta dt',\eea
\beq\label{16}
\Delta y(t)=\frac{1}{2}F\Delta\Gamma\int_0^t\sin2\theta
dt'+\int_0^t\xi_ydt'+v\int_0^t\sin\theta dt',\eeq with
averages given by
\bea\label{17}
\langle\Delta x\rangle_{\theta_0}=F\Gamma
t+v\tau_1\cos\theta_0+\frac{1}{2}F\Delta\Gamma\tau_4\cos2\theta_0,\nn \\
\langle\Delta y\rangle_{\theta_0}=v\tau_1\sin\theta_0+\frac{1}{2}F\Delta\Gamma\tau_4\sin2\theta_0.
\eea

The averages of the mean-squared displacements along the $x$ and $y$ axes are given by
\bea\label{18}
\langle\Delta x^2\rangle_{\theta_0}
=\frac{\tau_R}{24}F^2\Delta\Gamma^2[3(t-\tau_4)+(\tau_4-\tau_{16})\cos4\theta_0]\nn\\+\Delta D\tau_4\cos2\theta_0+F^2 \Gamma \Delta \Gamma t\tau_4\cos2\theta_0+2vF\Gamma t\tau_1\cos\theta_0\nn\\+2Dt+F^2\Gamma^2t^2+\frac{2}{3}\tau_Rv^2[3(t-\tau_1)+(\tau_1-\tau_4)\cos2\theta_0]\nn\\+\frac{1}{120}\tau_RvF\Delta\Gamma[40(4\tau_1-3te^{-D_Rt}-\tau_4)\cos\theta_0\nn\\+3(5\tau_1+8\tau_4-13\tau_9)\cos3\theta_0],\nn\\
\eea
\bea\label{19}
\langle\Delta
y^2\rangle_{\theta_0}
=\frac{\tau_R}{24}F^2\Delta\Gamma^2[3(t-\tau_4)-(\tau_4-\tau_{16})\cos4\theta_0]\nn \\
-\Delta D\tau_4\cos2\theta_0+\frac{2}{3}\tau_Rv^2[3(t-\tau_1)-(\tau_1-\tau_4)\cos2\theta_0]\nn\\
+\frac{1}{120}\tau_RvF\Delta\Gamma[40(4\tau_1-3te^{-D_Rt}-\tau_4)\cos\theta_0\nn \\
-3(5\tau_1+8\tau_4-13\tau_9)\cos3\theta_0]+2Dt,\nn \\
\eea
and the average of the cross-correlation between
displacements along the $x$ and $y$ axes is
\bea\label{20}
\langle\Delta x\Delta y\rangle_{\theta_0}
=\frac{\tau_R}{24}F^2\Delta\Gamma^2(\tau_4-\tau_{16})\sin4\theta_0+vF\Gamma t\tau_1\times\nn \\\sin\theta_0
+\frac{1}{2}F^2\Gamma\Delta\Gamma t\tau_4\sin2\theta_0+\frac{2}{3}\tau_Rv^2(\tau_1-\tau_4)\sin2\theta_0\nn\\
+\frac{1}{40}\tau_RvF\Delta\Gamma(5\tau_1+8\tau_4-13\tau_9)\sin3\theta_0,\nn\\
\eea
In our calculations we took into account that the terms odd in the
noise average to zero. Then the mean-squared displacement is

\bea\label{21}
\langle\Delta \textbf{r}^2\rangle =\frac{1}{2\pi}\int_0^{2\pi}d\theta_0\frac{\langle\Delta
y^2\rangle_{\theta_0}+\langle\Delta
x^2\rangle_{\theta_0}}{2t} \nn\\
 =4Dt+4\tau_Rv^2(t-\tau_1)+F^2\Gamma^2t^2\nn\\
 +\frac{1}{4}F^2\Delta\Gamma^2\tau_R(t-\tau_4),
\eea
and the long time $t\gg\tau_R$ diffusion tensor averaged over the initial angle,
$\theta_0$, yields the following experssion

\beq\label{22}
D_{ij}(t)=\frac{1}{2\pi}\int_0^{2\pi}d\theta_0\frac{\langle\Delta x_i\Delta
x_j\rangle_{\theta_0}}{2t}=D_{SF}\delta_{ij} \eeq 
 with
 
\bea\label{23}
D_{SF}=D+\tau_R
v^2+vF\Delta\Gamma\tau_R\nn\\+\frac{1}{16}F^2\Delta\Gamma^2\tau_R+\frac{1}{4}F^2\Gamma^2t.
\eea

\emph{\textbf{Self-propulsion vs. external potential field -}} We
saw that for the self-propelled particle the long time single
diffusion coefficient is enhanced according to $D_S=D+\tau_Rv^2$.
This enhancement is similar to that of a non self-propelled
ellipsoid due to external constant potential field
($D_F=D+\tau_R|\textbf{F}|^2\Delta\Gamma^2/16$ where $|\textbf{F}|$
is the force magnitude of the external potential
field~\cite{grima}~) since both of them are proportional to the
square of the force magnitude. However, the latter explicitly depends on the
asymmetry of the particle, $\Delta\Gamma$, but
the former does not. The origin of this discrepancy may be apparent from the fact that our model assumes that in addition to the external constant force the rod has a fixed speed $v$ directed along its long axis at all times while the former
model assumes the rod feels only an external constant force $\textbf{F}$ at all
times. Therefore similar to the hydrodynamic drag force, the
external potential field brings in the friction factor
between the two situations.

Also, a clear difference between the two situations is easily
understood by neglecting the noise and hence considering a deterministic
description. Then one can immediately calculate the velocity and
displacement in the lab frame as functions of time and the initial
angle, $\theta_0$, for given initial conditions. In the absence of
external potential field we have zero mean displacement and zero velocity when averaging over
$\theta_0$. However, in the presence of external potential field one obtains a mean velocity
$\Gamma F_x$ in the $x$ direction and $\Gamma F_y$ in the $y$
direction with corresponding mean displacements.

\section{Discussion}

In writing down the~Eqs.~\eqref{1}  and~\eqref{2} we have dropped the inertial
terms assuming the motion is overdamped, that is, the viscous terms
dominate the inertial ones. If we had kept those terms, they would
have generated homogenous part of the solutions which exponentially
decay with the characteristic \textit{relaxation times}, $m\times
\Gamma_{ij}$ for translational and $I\times \Gamma_R$ for rotational
motion. Here $m$ is the mass and $I$ is the moment of inertia about
the $z$ axis passing through the center of mass of the rod. Our results are,
therefore, not valid in time intervals comparable to the relaxation
times given the fact that recently developed experimentation
technics, namely, use of the Photonic Force Microscopy allows high
precision measurements in such small time intervals~\cite{micro}.

\emph{\textbf{Mean displacement -}} Eq.~\eqref{8} has limiting
forms of $|\langle \Delta \textbf{r}\rangle_{\theta_0}|=2v\tau_R$
for $t\gg\tau_R$ suggesting that at very long times the
self-propelled rod must behave as a spherical Brownian particle in
the region centered $2v\tau_R$ far from its initial position and
$|\langle \Delta \textbf{r}\rangle_{\theta_0}|=vt$ for $t\ll\tau_R$.
The latter is intuitive since at early times the motion is
anisotropic which is not the case for the long time result. To
give a full physical meaning to these results, we contrast the cases
when the self-propulsion (SP) is `turned on' and `off'. First thing
to note about these two situations is that both of them have the
same rotational time, $\tau_R$, since the net torque is zero in
both cases. Using this, we can picture what the rod would do before and after the crossover has
occurred when the SP is on and when it is off: (i) before the crossover ($t < \tau_R$), short time
anisotropy gets contribution from SP which causes the rod move faster
in a particluar direction than when SP is turned off. Therefore in the
anisotropic regime the rod displaces more due to SP than when SP
is turned off, (ii) after the crossover has occurred ($t
>  \tau_R$), in both cases the rod starts performing isotropic diffusion about
the position that it has arrived during  the time while $t <
\tau_R$. Therefore $2v\tau_R$ is the mean distance between the
\textit{centers of isotropic diffusion} domains of non self-propelled and
self-propelled rods.

Note that if we average over the initial orientation,
components of the mean displacement are zero and so is the mean
displacement. This is not in contradiction with what we have above.
Averaging over the initial angle is equivalent to averaging over the
number of non-interacting self-propelled rods that started up from
the same point in isotropic manner. At long times, the centers of
isotropic diffusion for these rods are distributed isotropically
about the starting point and hence have zero mean displacement. From
this point of view, we average the diffusion tensor over the initial
orientation to analyze transient behavior of diffusion in the system
of non-interacting such rods.

A particle undergoing classical Brownian motion takes the same
amount of time for going from a given initial point to a final point
and coming back. For our model this is not the case: The domain of
isotropic motion is centered $2v\tau_R$ far from an arbitrary
starting point implying that a particle would need more time
(equivalently less probability) for coming back. In fact, it has
been observed that when the \textit{pigment granules} are spreading
out from nucleus of the cell, they go straight through \textit{actin
filament} intersections until they get to the end of the filament,
but if they need to return to the nucleus, \textit{cargos} often
switch at actin filament intersections hence taking relatively
longer time to come back~\cite{clare}. This is in accordance with
the prediction of the model presented here.

\emph{\textbf{Mean-squared displacement -}} In Eq.~\eqref{11} the
second term is exactly the same as the one in refs~\cite{dunn, oth,
stok} obtained for the \textit{persistent random walk}~\cite{randi}
with persistence time $P=2\tau_R$ and the weak noise limit of the
expression obtained from a model of Brownian particles which are
pumped with energy by means of a non-linear friction function (upon
setting $\tau_R=v_0^2/4D$) in refs~\cite{ebeling0, ebeling}, while
the first term is the contribution due to \textit{classical Brownian
motion}. This is similar to that of the Levy-walk superdiffusive motion which can be decomposed into
thermal orientation fluctuations and an active motion of the rods
with a constant velocity along their long axis~\cite{dhar}.

It is known that at short times the cell appears to persist at the
direction of motion but at long times the pattern of movement is
similar to the classical Brownian motion~\cite{mark}. Because of
this, the persistent random walk model have been used to quantify
cell migrations~\cite{dee, dick, salt, dimil, glas, tan, alex,
fish}. The results presented here suggest that by including contribution due to
the classical Brownian motion as in Eq.~\eqref{11} one can describe
individual cell movements on a plane substrate more accurately than
it has been using the persistent random walk model.

Recently, an experiment has been carried out by Jonathan R. Howse,
\emph{et al}. who studied the motion of polystyrene spheres with a
diameter of 1.62 $\mu$m immersed in hydrogen peroxide and
half-coated with platinum~\cite{golestanian}. Due to the local
osmotic pressure gradient created by the asymmetric chemical
reaction such spheres were propelled. They were able to fit the
experimental data only with the short- and long-time limits of the
expression for their model, which is the same as Eq.~\eqref{11} if
we replace $\tau_R$ with $\tau_R/4$. But if we consider the
displacement components along the $x$ and $y$ axes separately, then
immediately we see a major difference between these two cases which
has been hidden in the expressions for the mean-squared
displacements. The first model consists of a self-propelled sphere
whose direction of motion is subject to diffusion and hence no
expression could contain shape asymmetry as it is zero. In contrast,
our model contains shape asymmetry, $\Delta D$, which makes it
distinct from any previously studied model when considering the $x$
and $y$ components of quantities of our interest, such as the
mean-squared displacement. We are able to recover all previously
obtained results upon setting $\Delta D=0$ in
our calculations.

In summery, here we present a simple model to describe the individual movements of rod-like living microorganisms on a plain. For that we studied the Brownian motion of an asymmetric rigid particle self-propelled at a fixed speed $v$ along its long axis in two dimensions.
The results for the case when such a particle is in external potential field are also presented. Some relevant biological systems are rod like living microorganisms such as the Gram-negative delta-proteobacterium \emph{Myxococcus xanthus}, Gram-negative bacterium \emph{Escherichia
coli} and \textit{Mitochondria}. Non-living self-propelled systems which resemble our model may be a
polar rod on a vibrating substrate~\cite{arshad}, self-driven
anisotropic colloidal particles and catalytically driven \textit{nanorods}~\cite{sven, bala}.
These results are expected to better quantify motion of such systems over the entire range of time.

\acknowledgments{I would like to thank M. Cristina Marchetti who posed the problem and supervised me during the earlier calculations. I also wish to thank Natsuhiko Yoshinaga, Arshad Kudrolli,
Eric Lauga, and Clare Yu for helpful discussion of the work during
the poster session of the I2CAM Workshop in Syracuse. }
\appendix

\section{}

Here all averages $\langle...\rangle$ are taken at a fixed initial
angle, $\theta_0$, and $\theta_1\equiv\theta(t_2)$ and
$\theta_1\equiv\theta(t_2)$.

We start by writing
\bea\label{a1}
\langle n_i(t_1)
n_j(t_2)\rangle=\left(
                             \begin{array}{cc}
                               \langle\cos\theta_1\cos\theta_2\rangle & \langle\cos\theta_1\sin\theta_2\rangle \\
                               \langle\sin\theta_1\cos\theta_2\rangle & \langle\sin\theta_1\sin\theta_2\rangle \nn\\
                             \end{array}
                           \right),\\
\eea and
\bea\label{a2}
\cos\theta_1\cos\theta_2=\frac{1}{2}[\cos(\theta_1-\theta_2)+\cos(\theta_1+\theta_2)],\nn \\
\cos\theta_1\sin\theta_2=\frac{1}{2}[\sin(\theta_1+\theta_2)-\sin(\theta_1-\theta_2)],\nn \\
\sin\theta_1\cos\theta_2=\frac{1}{2}[\sin(\theta_1-\theta_2)+\sin(\theta_1+\theta_2)],\nn \\
\sin\theta_1\sin\theta_2=\frac{1}{2}[\cos(\theta_1-\theta_2)-\cos(\theta_1+\theta_2)].\nn \\
\eea To calculate the averages we use Eq.~\eqref{6} and the fact
that $\cos\theta=\Re e^{i\theta}$ and $\sin \theta =\Im e^{i\theta}$
where $\Re$ and $\Im$ stand for the real and imaginary parts
respectively. Then
\bea\label{a3}
2\langle\cos\theta(t_1)\cos\theta(t_2)\rangle=e^{-D_R[t_1+t_2-2\min(t_1,t_2)]}\nn\\+\cos2\theta_0 e^{-D_R[t_1+t_2+2\min(t_1,t_2)]},\nn \\
2\langle\sin\theta(t_1)\sin\theta(t_2)\rangle=e^{-D_R[t_1+t_2-2\min(t_1,t_2)]}\nn\\-\cos2\theta_0 e^{-D_R[t_1+t_2+2\min(t_1,t_2)]},\nn \\
2\langle\cos\theta(t_1)\sin\theta(t_2)\rangle=\sin2\theta_0 e^{-D_R[t_1+t_2+2\min(t_1,t_2)]},\nn \\
2\langle\sin\theta(t_1)\cos\theta(t_2)\rangle=\sin2\theta_0e^{-D_R[t_1+t_2+2\min(t_1,t_2)]}.\nn\\
\eea
If we put Eq.~\eqref{a3} into Eq.~\eqref{a1}, we obtain
\bea\label{a4}
2\langle n_i(t_1)n_j(t_2)\rangle_{\theta_0}=\delta_{ij}e^{-D_R[t_1+t_2-2\min(t_1,t_2)]}\nn\\+M_{ij}(\theta_0)e^{-D_R[t_1+t_2+2\min(t_1,t_2)]}\eea
Now we need to integrate Eq.~\eqref{a4} twice. To do that we use
the following formulae
\beq\label{a5}
\int_0^{t}d t_1\int_0^{t}d t_2
...=\underbrace{\int_0^{t}d t_1\int_0^{t_1}d t_2 ...}_{\hbox{$t_1>t_2$}}+\underbrace{\int_0^{t}d
t_2\int_0^{t_2}d t_1 ...}_{\hbox{$t_1<t_2$}}, \eeq 
which can easily be proved using geometrical considerations. Then we have
\bea\label{a6} 
\int_0^{t}dt_1\int_0^{t}d t_2e^{-D_R[t_1+t_2+2\min(t_1,t_2)]}\nn\\=2\int_0^{t}d
t_1\int_0^{t_1}d t_2e^{-D_R[t_1+3t_2]}=\frac{4\tau_R}{3}[\tau_1(t)-\tau_4(t)],\nn \\
\int_0^{t}d t_1\int_0^{t}dt_2e^{-D_R[t_1+t_2-2\min(t_1,t_2)]}\nn\\=2\int_0^{t}d
t_1\int_0^{t_1}d t_2 e^{-D_R[t_1-t_2]}=4\tau_R[t-\tau_1(t)].\nn\\
\eea

Finally, if we put Eqs.~\eqref{a4}, and ~\eqref{a6} together, we
obtain
\bea\label{a7}
v^2\int_0^{t}d t_1\int_0^{t}d t_2\langle
n_i(t_1) n_j(t_2)\rangle_{\theta_0}\nn\\=
2\tau_Rv^2[t-\tau_1]\delta_{ij}
+\frac{2}{3}\tau_Rv^2[\tau_1-\tau_4]M_{ij}(\theta_0).
\eea

\end{document}